\documentclass[
reprint, 
superscriptaddress,
amsmath,
amssymb,
aps,
prl,
longbibliography,
]{revtex4-2}

\usepackage{graphicx}
\usepackage{dcolumn}
\usepackage{physics}
\usepackage{gensymb}
\usepackage{bm}
\usepackage{xcolor}
\usepackage{footmisc}

\begin{document}

\preprint{APS/123-QED}

\title{State-dependent phonon-limited spin relaxation of nitrogen-vacancy centers}

\author{M.~C.~Cambria}
\thanks{These authors contributed equally.}

\author{A.~Gardill}
\thanks{These authors contributed equally.}

\author{Y.~Li}
 
\affiliation{Department of Physics, University of Wisconsin, Madison, Wisconsin 53706, USA}

\author{A.~Norambuena}
 
\affiliation{Centro de Investigaci\'on DAiTA Lab, Facultad de Estudios Interdisciplinarios, Universidad Mayor, Santiago, Chile}

\author{J.~R.~Maze}
 
\affiliation{Instituto de F\'isica, Pontificia Universidad Cat\'olica de Chile, Casilla 306, Santiago, Chile}

\affiliation{Centro de Investigaci\'on en Nanotecnolog\'ia y Materiales Avanzados, Pontificia Universidad Cat\'olica de Chile, Santiago, Chile}

\author{S.~Kolkowitz}
\email{kolkowitz@wisc.edu}
 
\affiliation{Department of Physics, University of Wisconsin, Madison, Wisconsin 53706, USA}

\date{\today}
\begin{abstract}
Understanding the limits to the spin-coherence of the nitrogen-vacancy (NV) center in diamond 
is vital to realizing the full potential of this quantum system. 
We show that relaxation on the \(\ket{m_{s}=-1} \leftrightarrow \ket{m_{s}=+1}\) transition occurs approximately twice as fast as relaxation on the \(\ket{m_{s}=0} \leftrightarrow \ket{m_{s}=\pm 1}\) transitions under ambient conditions in native NVs in high-purity bulk diamond. The rates we observe are independent of NV concentration over four orders of magnitude, indicating they are limited by spin-phonon interactions. 
We find that the maximum theoretically achievable coherence time for an NV at 295 K is limited to 6.8(2) ms. Finally, we present a theoretical analysis of our results that suggests Orbach-like relaxation from quasilocalized phonons or contributions due to higher-order terms in the spin-phonon Hamiltonian are the dominant mechanism behind \(\ket{m_{s}=-1} \leftrightarrow \ket{m_{s}=+1}\) relaxation, motivating future measurements of the temperature dependence of 
this relaxation rate.
\end{abstract}

\maketitle

\section{Introduction}

The nitrogen-vacancy (NV) center in diamond is a solid-state quantum system that has attracted interest for a range of potential applications, such as quantum information processing \cite{Neumann2010, Dolde2012} and sensing of the local magnetic \cite{Rondin2014}, electric \cite{Dolde2011, Mittiga2018}, strain \cite{Ovartchaiyapong2014,Teissier2014}, and thermal \cite{Kucsko2013, Toyli2013} environment. Critical to the performance of the NV center in many of these applications is its long electronic spin coherence time at room temperature. Coherence times of approximately 2--3 ms have previously been demonstrated in bulk diamond at room temperature \cite{Naydenov2011,bar_gill2013}. However, despite the use of sophisticated dynamical decoupling sequences and high-purity diamond samples, these values remain far short of the measured lifetime of the \(\ket{m_{s}=0}\) state (often called the $T_1$ time), prompting consideration of the validity of treating the NV as a two-level system \cite{bar_gill2013}. Furthermore, the phonon-limited lifetimes of the \(\ket{m_{s}= \pm 1}\) states have not previously been investigated, leaving an incomplete picture of the NV's spin-lattice relaxation dynamics. As a result, the ultimate limit to the NV electronic spin coherence time at room temperature has remained unclear.

Numerous experimental measurements of the lifetime of the \(\ket{m_{s}=0}\) state in a range of samples with varying defect concentrations have indicated that spin-lattice interactions are the primary mechanism for spin relaxation at room temperature in high-purity bulk diamond samples \cite{Jarmola2012}. 
Specifically, two-phonon Raman transitions with major contributions from acoustic and quasilocalized phonons are understood to account for the experimentally observed temperature dependence of the rate of relaxation out of \(\ket{m_{s}=0}\) near room temperature \cite{Norambuena2018, Redman1991, Takahashi2008, Jarmola2012}. 
Prior work on phonon-induced relaxation has focused on the lifetime of the \(\ket{m_{s}=0}\) state, however, neglecting possible relaxation on the \(\ket{m_{s} = -1} \leftrightarrow \ket{m_{s} = +1}\) transition. Recent work with near-surface NVs has demonstrated fast relaxation on the \(\ket{m_{s} = -1} \leftrightarrow \ket{m_{s} = +1}\) transition, with relaxation rates of up to approximately \(2 \times 10^{3} \text{ s}^{-1}\) and \(2 \times 10^{5} \text{ s}^{-1}\) observed in shallow NVs \cite{Myers2017} and NVs in nanodiamond \cite{Gardill2020} respectively. In both contexts this relaxation is attributed to electric field noise emanating from the diamond surface. Deep in bulk diamond, far less electric field noise can be expected, leaving open the question of the phonon-induced relaxation rate of the \(\ket{m_{s} = -1} \leftrightarrow \ket{m_{s} = +1}\) transition as well as the role of other impurities in the diamond lattice.

In this Letter we present experimental results indicating that phonon-induced relaxation on the \(\ket{m_{s}=-1} \leftrightarrow \ket{m_{s}=+1}\) transition occurs at roughly twice the rate of relaxation on the \(\ket{m_{s}=0} \leftrightarrow \ket{m_{s}= \pm 1}\) transition in ambient conditions, setting an ultimate limit on the maximum achievable NV electronic spin coherence time of 6.8(2) ms for a superposition of \(\ket{m_{s} = 0}\) and \(\ket{m_{s} = \pm1}\) at room temperature (295 K) in high-purity bulk diamond. We discuss the importance of considering the full picture of the NV qutrit, rather than the simplified qubit model. In addition, we show that the standard theoretical model of two-acoustic-phonon Raman processes invoked to explain NV spin-lattice relaxation, which is thought to significantly contribute to relaxation on the \(\ket{m_{s}=0} \leftrightarrow \ket{m_{s}= \pm 1}\) transition, predicts a rate of zero for the \(\ket{m_{s}=-1} \leftrightarrow \ket{m_{s}=+1}\) transition as a consequence of time-reversal symmetry. This suggests that two-quasilocalized-phonon Orbach-like processes or processes involving a spin-phonon Hamiltonian to second order in the atomic displacements are a dominant source of spin-phonon relaxation on the \(\ket{m_{s}=-1} \leftrightarrow \ket{m_{s}=+1}\) transition at room temperature. 
Future studies of relaxation on the \(\ket{m_{s}=-1} \leftrightarrow \ket{m_{s}=+1}\) transition as a function of temperature could resolve the relative contributions of these different mechanisms.

As a matter of terminology, we note that the existing literature refers to the same transitions with various names. 
The \(\ket{m_{s}=-1} \leftrightarrow \ket{m_{s}=+1}\) and \(\ket{m_{s}=0} \leftrightarrow \ket{m_{s}= \pm 1}\) transitions have been referred to as double-quantum and single-quantum transitions respectively, but for mixed eigenstates these terms may be misleading. The names of the transitions were generalized to account for this in Ref.~\cite{Gardill2020}, and we will adopt that naming scheme here. Specifically, with \(\ket{H;i}\) denoting the energy eigenstate with majority component \(\ket{m_{s}=i}\), we call the \(\ket{H;-1} \leftrightarrow \ket{H;+1}\) and \(\ket{H;0} \leftrightarrow \ket{H;\pm1}\) transitions the qutrit and qubit transitions respectively. In referring to the eigenstates going forward, we will omit the \(H\) identifier from kets, referring to \(\ket{H;i}\) simply as \(\ket{i}\). 

\begin{figure}[b]
\vspace{-0.3cm}
\includegraphics[width=0.48\textwidth]{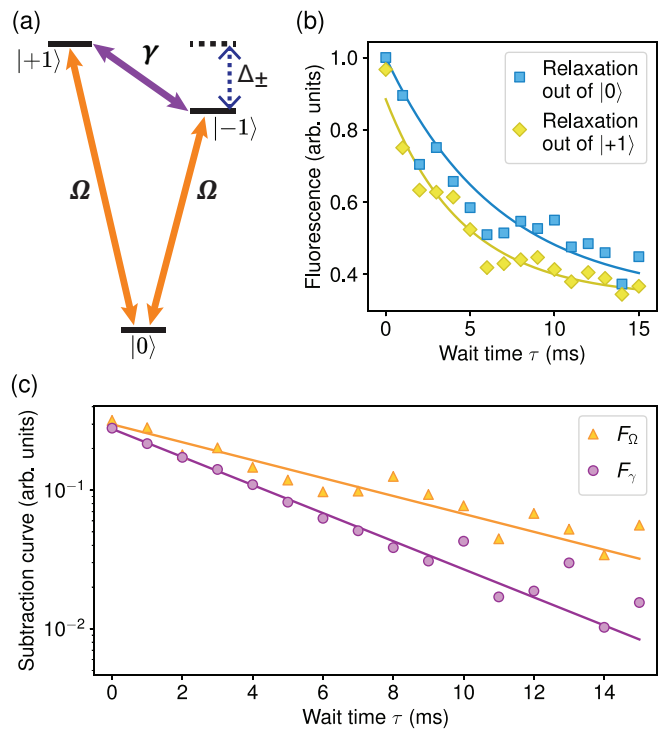}
\caption{\label{fig:relaxation} State-dependent spin relaxation rates of nitrogen-vacancy centers. (a) Ground state level-structure of the NV\(^{-}\) electronic spin. Qubit transitions between \(\ket{0}\) and \(\ket{\pm1}\) occur at rate \(\Omega\), while qutrit transitions between \(\ket{+1}\) and \(\ket{-1}\) occur at rate \(\gamma\). An applied magnetic field lifts the \(\ket{\pm1}\) degeneracy by the splitting \(\Delta_{\pm}\). (b) Measurement of state-dependent decay curves. Relaxation out of \(\ket{0}\) (cyan squares) is described by a single exponential with decay rate \(3\Omega\) (cyan curve). Relaxation out of \(\ket{+1}\) (yellow diamonds) is described by a biexponential decay that depends on both \(\Omega\) and \(\gamma\) (yellow curve). Solid lines are fits to population dynamics equations of the ground state \cite{Gardill2020}. (The deviation of the yellow curve from 1 at \(\tau=0\) is due to \(\pi\)-pulse infidelity.) For data shown, \(\Omega = 56(3) \text{ s}^{-1}\), \(\gamma = 132(11) \text{ s}^{-1}\). (c) Semilog plot of subtracted population curves, \(F_{\Omega}\) and \(F_{\gamma}\) in Eqs.~\ref{eqs:FFuns1} and \ref{eqs:FFuns2}, used to extract $\Omega$ and $\gamma$ respectively. Displayed data is from the same measurement series as that shown in (b).}
\end{figure}

\section{Methods}

Experiments were conducted under ambient conditions using a homebuilt confocal microscope with a temperature stable to \(295 \text{ K}\) within \(\pm 1 \text{ K}\). Applied magnetic fields were typically \(< 100 \text{ G}\), with few measurements conducted at higher fields. (Table S1 of the Supplemental Material \cite{SupplementalMaterials} contains a complete summary of the data collected for this paper, including applied magnetic fields.) The components of the applied magnetic field are defined as on- and off-axis with respect to the NV's spatial axis of symmetry. 
Spin polarization and readout were accomplished with approximately 1 mW of 532-nm light. 

To extract the qubit and qutrit relaxation rates, defined as $\Omega$ and $\gamma$ respectively as in Fig.~\ref{fig:relaxation}(a), we use state-selective \(\pi\)-pulses to measure the population decay into and out of each spin state individually, as described in \cite{Myers2017, Gardill2020}. Under a classical three-level population model, subtraction of these curves yields single-exponential decays from which we isolate the relaxation rates for each transition. In particular, with \(P_{i,j}(\tau)\) denoting the population in \(\ket{j}\) at time \(\tau\) after initialization in \(\ket{i}\), we determine $\Omega$ and $\gamma$ using the differences defined
 
\addtolength{\jot}{7pt}\begin{gather}
    \label{eqs:FFuns1}
    F_\Omega(\tau) = P_{0,0}(\tau) - P_{0,+1}(\tau) = a e^{-3\Omega\tau}, \\
    F_\gamma(\tau) =  P_{+1,+1}(\tau) - P_{+1,-1}(\tau) = a e^{-(2\gamma + \Omega)\tau},
    \label{eqs:FFuns2}
\end{gather}\addtolength{\jot}{-7pt}\noindent

\noindent where $a$ is the fluorescence contrast between the two subtracted data sets. Fig.~\ref{fig:relaxation}(c) shows a representative result of the single-exponential decay curves \(F_\Omega\) and \(F_\gamma\). In ensemble measurements, where multiple NVs with different orientations contribute to the measured signal, the applied magnetic field and the subtraction of population curves isolates the dynamics of the single NV orientation resonantly addressed by the microwaves from among the four orientations allowed by the diamond's tetrahedral lattice. 

\begin{figure}[b]
\vspace{-0.3cm}
\includegraphics[width=0.48\textwidth]{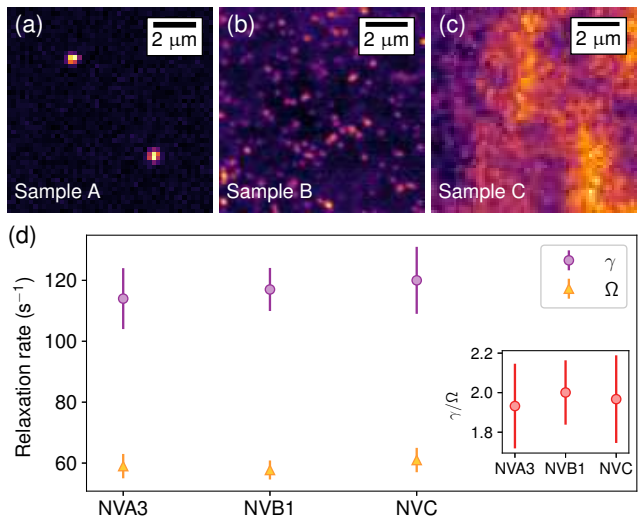}
\caption{\label{fig:samples}
Rates \(\gamma\) and \(\Omega\) for samples with differing defect concentration. (a-c) Confocal microscope images of samples A, B, and C, displaying NV concentrations ranging from approximately \(10^{-5}\) ppb to \(10^{-1}\) ppb. (d) Average measured rates \(\gamma\) and \(\Omega\) at small off-axis magnetic fields in samples A, B, and C. On-axis magnetic fields are in the range \(20-60 \text{ G}\). Measurements of both \(\gamma\) and \(\Omega\) are equivalent to within error in all three samples, indicating no dependence on defect concentration in the measured regime. 
Inset: The ratio \(\gamma / \Omega \approx 2\) for each pair of points shown in the main plot. Error bars are one standard error.}
\end{figure}

Relaxation measurements were performed on native NVs in three chemical-vapor-deposition (CVD) grown bulk diamond samples from different growers with different native NV concentrations. Confocal microscope images of the three samples can be seen in Fig.~\ref{fig:samples}(a-c). The NV concentrations of samples A and B were estimated by NV counting. The NV concentration of sample C was estimated by comparing the typical fluorescence of the sample to the fluorescence of the isolated single NVs visible in sample A. Sample A, from Element Six, has the lowest NV concentration at approximately \(10^{-5}\) ppb. Sample B, from Diamond Elements, has an NV concentration of approximately \(10^{-3}\) ppb. The NV concentrations of samples A and B are low enough to allow for measurements to be performed on single NVs using confocal microscopy. The NV concentration of Sample C, from Chenguang Machinery \& Electric Equipment Company, is high enough to prohibit measurement on single NVs in our confocal microscope, at approximately \(0.1\) ppb. For this sample we performed measurements on the ensemble of NVs within the confocal volume of the microscope. In the data, NVs are indexed first by sample and then numerically to distinguish single NVs where appropriate. Using the ratio of approximately 300 substitutional nitrogen defects per NV\(^{-}\) determined in \cite{Edmonds2012} for native defects in CVD-grown (100)-oriented samples, we estimate the substitutional nitrogen concentrations to be approximately \(3 \times 10^{-3}\) ppb, \(3 \times 10^{-1}\) ppb, and \(3 \times 10\) ppb in samples A, B, and C respectively.

\section{Results}

The measured relaxation rates $\gamma$ and $\Omega$ under small off-axis magnetic fields (quantitatively, \(B_{\perp} < 1\) G) for native NVs deep in the bulk in the three samples are shown in Fig.~\ref{fig:samples}(d). We find that neither \(\gamma\) nor \(\Omega\) depends on the NV concentration over a range covering four orders of magnitude from approximately \(10^{-5}\text{--}10^{-1}\) ppb. This indicates that the relaxation we observe is not the result of interactions between defects. In the small off-axis magnetic field regime, we find that relaxation on the qutrit transition is approximately twice as fast as relaxation on the qubit transition. Specifically, we measure the weighted averages \(\gamma=117(5) \text{ s}^{-1}\) and \(\Omega=59(2) \text{ s}^{-1}\). Importantly, this indicates that significant dynamics are missed if the NV is treated as a qubit by considering only one of the \(\ket{0}\) and \(\ket{\pm1}\) subspaces; population leakage to the third state actually occurs faster than relaxation between the states intended to comprise the qubit.

The standard $T_1$ relaxation experiment that is most commonly used with NVs consists of measuring the lifetime of the $\ket{0}$ state. The $T_1$ times quoted in the literature based on these measurements can be related to our values of $\Omega$ by defining $T_1^{(0)} = 1/3\Omega$ \cite{Myers2017}. As two examples, the \(T_{1}^{(0)}\) times reported from both Naydenov \emph{et al.} \cite{Naydenov2011} and Bar-Gill \emph{et al.} \cite{bar_gill2013} in isotopically pure diamond samples at room temperature corresponds to $\Omega =$ 56(4) \(\text{s}^{-1}\), in agreement with our measurements. To our knowledge, the qutrit relaxation rate $\gamma$ has not been systematically studied for NVs deep within bulk diamond samples. 

We next examine the limit to coherence time \(T_{2}\) imposed by the qubit and qutrit relaxation rates. We introduce, $T_{\text{2,max}}$, the maximum theoretically achievable coherence time due to incoherent relaxation between states. For a qubit basis formed of \(\ket{0}\) and either \(\ket{\pm1}\) \cite{Myers2017, Gardill2020},
\begin{equation}\label{t2_max_qubit}
    T_{\text{2,max}} = \frac{2}{3\Omega + \gamma}.
\end{equation}
For the values of $\Omega$ and $\gamma$ measured at small off-axis magnetic fields, we find $T_{\text{2,max}} = 6.8(2)$ ms. This value represents the maximum achievable coherence time at room temperature for NVs deep in high-purity bulk diamond samples if all pure dephasing sources are eliminated. As a caveat, we note that \(T_{\text{2,max}}\) could in principle be extended by engineering the vibrational modes of the lattice to suppress relaxation due to spin-lattice interactions.

As an example of the implications of this limit, $T_{\text{2,max}}$ determines the minimum theoretically achievable sensitivity of a room temperature single NV electronic spin to magnetic fields. If we assume $100\%$ charge state initialization and spin state polarization as well as zero dead time between measurements, the ultimate limit is given by \cite{Maze2008, Taylor2008}:
\begin{equation}\label{B_sensitivity_max}
    \delta B_{\text{min}} = \frac{\hbar}{g \mu_{\text{B}} C \sqrt{T_{\text{2,max}} T}} 
\end{equation}
where $T$ is the total averaging time, $\hbar$ is the reduced Planck constant, $g \approx 2$ is the electronic Land\'{e} $g$-factor, $\mu_{\text{B}}$ is the Bohr magneton, and \(C \leq 1\) describes the measurement efficiency. In our experiment \(C \approx 0.02\). Using the value of the qubit $T_{\text{2,max}}$ determined by our measurement of \(\gamma\) and \(\Omega\), we calculate $\delta B_{\text{min}} = 69(1)~\text{pT}/\sqrt{\text{Hz}}$ for the idealized case \(C=1\). As a practical example, this corresponds to the ability to detect the magnetic field from a single nuclear spin 30 nm away from the NV after 1 s of measurement time.

There has been considerable interest in using coherent superpositions of the $\ket{+1}$ and $\ket{-1}$ states for magnetic field sensing. This ``double-quantum'' basis eliminates noise from temperature drift and fluctuations of the electric and scaled-strain field along the NV axis while also doubling the frequency shift due to the magnetic field along the NV axis \cite{Mamin2014, Kim2015, Bauch2018, Bluvstein2019}. However, a sensor in this basis is also more susceptible to the faster qutrit decay $\gamma$. We next investigate the sensitivity of an ideal NV sensor making use of this basis, where the maximum theoretically achievable coherence time is \cite{Myers2017}
\begin{equation}\label{t2_max_qutrit}
    T_{\text{2,max}} = \frac{1}{\Omega + \gamma}.
\end{equation}
Using the same values of $\Omega$ and $\gamma$ as before, we find the maximum theoretically achievable coherence time for the double-quantum basis is $T_{\text{2,max}} = 5.7(2)$ ms, which is approximately 1 ms shorter than that in the ``single-quantum'' basis. This corresponds to a magnetic field sensitivity of $\delta B_{\text{min}} = 38(1)~\text{pT}/\sqrt{\text{Hz}}$, indicating that despite the shorter coherence time, an overall sensitivity advantage is conferred due to the factor of two increase in signal.
We note that if we set \(\gamma = 2\Omega\) in accordance with our measurements for small off-axis magnetic fields, we find \(T_{\text{2,max}} = 2/(5\Omega)\) in the single-quantum basis and \(T_{\text{2,max}} = 1/(3\Omega)\) in the double-quantum basis.

\begin{figure}[t]
\includegraphics[width=0.48\textwidth]{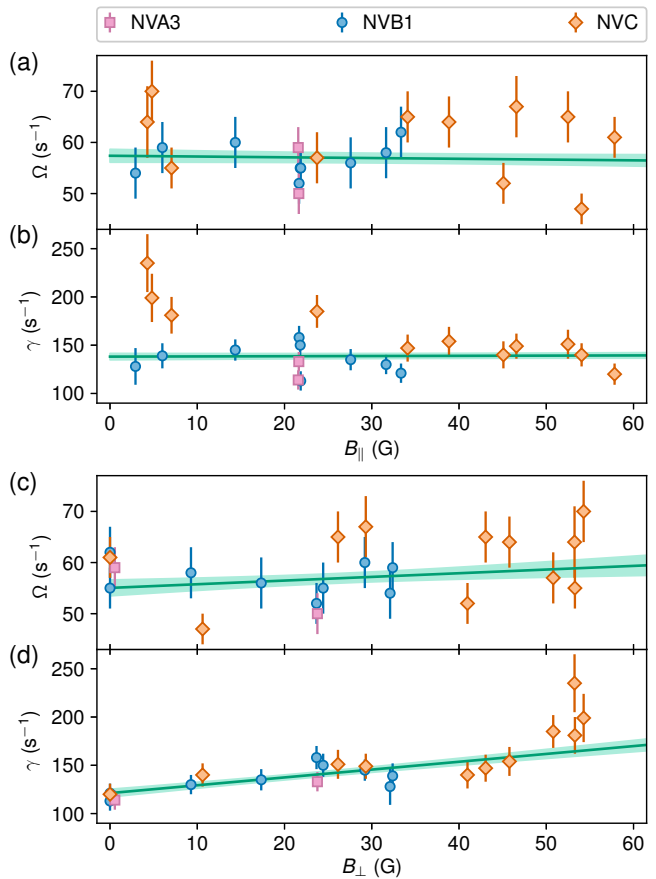}
\caption{\label{fig:rates}Measurements of \(\Omega\) and \(\gamma\) as functions of \(B_{\parallel}\) and \(B_{\perp}\), the on-axis and off-axis components of the applied magnetic field respectively. 
To allow for visual separation between the points in the figure, only measurements made at applied magnetic field magnitudes under 65 G are shown. (See \cite{SupplementalMaterials} for an extended version of this figure and a table of complete measurements.) Green lines are weighted linear regressions intended to show correlations. The regressions account for all data, including that not shown in the plots. Shaded regions demonstrate one standard error. 
The regressions evidence a small correlation of \(\gamma\) with \(B_{\perp}\); the other slopes are zero within two standard errors. 
Note that the same values of \(\Omega\) are plotted in (a) and (c). Likewise, the same values of \(\gamma\) are plotted in (b) and (d).
} 
\end{figure}

Figure~\ref{fig:rates} shows measurements of \(\Omega\) and \(\gamma\) for varying applied magnetic field magnitudes and orientations plotted as functions of $B_\parallel$ and $B_\perp$, the on-axis and off-axis components of the magnetic field respectively. 
Linear regressions are shown to demonstrate correlations between the rates and the field components. As functions of \(B_{\parallel}\), \(\Omega\) and \(\gamma\) display slopes of \(-0.01(2) \text{ s}^{-1}\text{/G}\) and \(0.02(5) \text{ s}^{-1}\text{/G}\) respectively. As functions of \(B_{\perp}\), \(\Omega\) and \(\gamma\) display slopes of \(0.07(5) \text{ s}^{-1}\text{/G}\) and \(0.81(14) \text{ s}^{-1}\text{/G}\) respectively. 
Interestingly, the slope of \(\gamma\) as a function of \(B_{\perp}\) (Fig.~\ref{fig:rates}(d)) indicates there is a significant but small correlation between the two variables, with \(\gamma\) accruing an approximate \(1\%\) increase per G of applied field. The analysis of directional correlations shown here provides additional information which is not typically captured in studies of phonon-limited relaxation that focus on temperature dependence. 
Consideration of such properties may help uncover and explain rich spin-lattice dynamics in NVs. 
In the discussion of maximum coherence times above and for the remainder of this work we focus on measurements conducted under small off-axis magnetic fields because superior phonon-limited coherence times are offered in this limit and because NV applications are commonly conducted with magnetic fields applied along the NV axis. 

\section{Discussion}

We now turn to a discussion of the origins of relaxation on the qutrit transition. 
In high-purity bulk diamond at room temperature, NV spin relaxation is attributed to spin-lattice interactions \cite{Redman1991, Takahashi2008, Jarmola2012, Norambuena2018}. 
The analysis by Jarmola \textit{et al.} \cite{Jarmola2012} found that a sum of the form 
\begin{equation}\label{empirical_fit}
    \frac{1}{T_{1}^{(0)}} = A_{1}(S) + \frac{A_{2}}{\exp(\Delta/k_{\text{B}}T)-1} + A_{3}T^{5}
\end{equation}
provides a good fit to experimental measurements of the rate of relaxation out of \(\ket{0}\) as a function of temperature in a range of samples. The fit parameter \(A_{1}(S)\) is a sample-dependent constant, while \(A_{2} = 2.1(6) \times 10^{3} \text{ s}^{-1}\) and \(A_{3} = 2.2(5) \times 10^{-11} \text{ s}^{-1}\) are coefficients weighting the relative contributions of the exponential and \(T^{5}\) terms. The parameter \(\Delta = 73(4) \text{ meV}\) is an empirical activation energy. Only the latter two terms in Eq.~\ref{empirical_fit} provide significant contributions at temperatures \(T \gtrsim 50 \text{ K}\), in agreement with the earlier analysis of Ref.~\cite{Redman1991} and the work of Ref.~\cite{Norambuena2018}, which considers a wider variety of low-temperature relaxation phenomena. While past results uniformly emphasize the significance of the exponential and \(T^{5}\) terms at high temperatures, claims regarding which of these terms is dominant at room temperature have been mixed. Using the numbers from Ref.~\cite{Jarmola2012}, we calculate that the exponential term contributes approximately \(120 \text{ s}^{-1}\) and the \(T^{5}\) term contributes approximately \(50 \text{ s}^{-1}\) to \(1/T_{1}^{(0)} \approx 170 \text{ s}^{-1}\) at \(T = 295 \text{ K}\). 

The mechanism behind the exponential term has been identified as a two-quasilocalized-phonon Orbach-like process. The empirical activation energy associated with this term is roughly consistent with \textit{ab initio} calculations that have revealed a band of quasilocalized phonon modes centered at 65 meV \cite{Zhang2011, Gali2011, Alkauskas2014}. The \(T^{5}\) term is attributed to a two-acoustic-phonon Raman process.
The theoretical basis for this identification comes from Ref.~\cite{Walker1968}, in which Walker demonstrates a two-acoustic-phonon Raman process leading to transition rates with a thermal scaling of \(T^{5}\). 

\begin{figure}[t]
\includegraphics[width=0.48\textwidth]{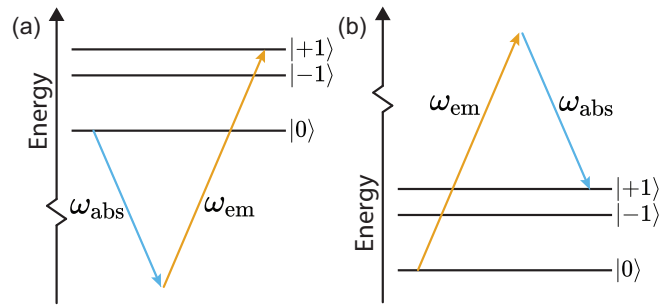}
\caption{\label{fig:phonons} Two examples of two-phonon processes driving the qubit transition. In both processes, a phonon of energy \(\hbar \omega_{\text{em}}\) is emitted and a phonon of energy \(\hbar \omega_{\text{abs}}\) is absorbed. Importantly, the emission-first process (a) is associated with a different suppressive factor than the absorption-first process (b) in perturbation theory. 
}
\end{figure}

The Supplemental Material \cite{SupplementalMaterials} contains derivations of the two-acoustic-phonon Raman process in Ref.~\cite{Walker1968} and a phenomenological model of a two-quasilocalized-phonon Orbach-like process. Both derivations are based on a spin-lattice Hamiltonian taken to first order in the atomic displacements and applied to second-order in time-dependent perturbation theory. The processes may proceed by either absorption-followed-by-emission or emission-followed-by-absorption, as shown in Fig.~\ref{fig:phonons}. Due to the time-reversal symmetry of the spin-lattice Hamiltonian, a generic two-phonon Raman process to first-order in the atomic displacements and second order in perturbation theory only provides a nonzero contribution to \(\gamma\) by the inequality of these two variants of the process. 
Because this restriction is inconsistent with the assumptions in Ref.~\cite{Walker1968} leading to a \(T^{5}\) thermal scaling of a two-acoustic-phonon Raman process (which has previously been invoked to explain the \(T^{5}\) term in Eq.~\ref{empirical_fit}), we conclude that this theoretical model cannot explain our observation of significant relaxation on the qutrit transition.

In contrast, a phenomenological model of the effect of two quasilocalized phonons results in an Orbach-like process (the exponential term in Eq.~\ref{empirical_fit}) 
that contributes nonzero $\gamma$ and $\Omega$ relaxation rates. These observations indicate that two-quasilocalized-phonon Orbach-like processes may be the dominant mechanism for relaxation on the qutrit transition at room temperature. Along with the observation that \(\gamma > \Omega\), this suggests that contributions from quasilocalized phonons may also be dominant for relaxation on the qubit transitions. While the simplicity of the ratio \(\gamma/\Omega \approx 2\) is suggestive, we are unable to say conclusively whether this is a coincidence arising from various contributions to the two rates, or whether it is because a single mechanism dominates both the qutrit and qubit relaxation rates and naturally gives rise to the ratio between the rates. 

Other possible contributions from two-phonon processes arise due to the action of a spin-lattice Hamiltonian taken to second order in the atomic displacements and applied to first order in perturbation theory \cite{Van1954}. These contributions are not typically considered for the NV; where they have been considered, symmetry arguments such as the one discussed here have not been applied \cite{Norambuena2018}. We include a brief discussion of these contributions in the Supplemental Material \cite{SupplementalMaterials}. We leave for future work a complete investigation of the effect of the second-order terms in the spin-lattice Hamiltonian in light of symmetry considerations. Ultimately, the difference in expected temperature scalings of the relaxation rates induced by the various processes provides a way to potentially disentangle their contributions, motivating future measurements of $\gamma$, and the ratio $\gamma/\Omega$, as a function of temperature. \textit{Ab initio} calculations of the coupling constants describing the spin-lattice Hamiltonian may also be critical in explaining the ratio \(\gamma/\Omega \approx 2\) \cite{Gugler2018}.

\section{Conclusions}

We have presented observations of spin state-dependent relaxation rates under ambient conditions in high-purity bulk diamond samples. We have measured the rate of phonon-limited relaxation on the qutrit transition, $\gamma$, at room temperature. Our measurements provide a complete experimental picture of the three-level relaxation dynamics in high-purity diamond at room temperature, indicating $T_\text{2,max} = 6.8(2)$ ms as the limit on the maximum theoretically achievable coherence time of an NV electronic spin at 295 K. Furthermore, we have shown that the standard model of two-acoustic-phonon Raman processes predicts no relaxation on the qutrit transition, motivating measurement of the temperature dependence of \(\gamma\) and providing evidence that quasilocalized phonons or contributions from higher order terms in the spin-phonon Hamiltonian provide dominant contributions to spin-lattice relaxation on the qutrit transition at room temperature. 

\section{Acknowledgments}

\noindent The authors thank Jeff Thompson, Nathalie de Leon, Gregory Fuchs, Dolev Bluvstein, Nir Bar-Gill, and Norman Yao for enlightening discussions, helpful insights, and comments on the manuscript. Experimental work, data analysis, and theoretical efforts conducted at UW--Madison were supported by the U.S. Department of Energy (DOE), Office of Science, Basic Energy Sciences (BES) under Award \#DE-SC0020313. Theoretical contributions conducted at Pontificia Universidad Cat\'olica de Chile by J.~R.~M.~ and A.~N.~ were supported by ANID Fondecyt 1180673 and ANID PIA ACT192023. A.~N.~ acknowledges financial support from Universidad Mayor through the Postdoctoral Fellowship. A.~G.~acknowledges support from the Department of Defense through the National Defense Science and Engineering Graduate Fellowship (NDSEG) program.

\bibliography{MainReferences.bib}

\end{document}


\preprint{APS/123-QED}

\title{Supplemental Materials for ``State-dependent phonon-limited spin relaxation of nitrogen-vacancy centers''}

\author{M.~C.~Cambria}
\thanks{These authors contributed equally.}

\author{A.~Gardill}
\thanks{These authors contributed equally.}

\author{Y.~Li}
 
\affiliation{Department of Physics, University of Wisconsin, Madison, Wisconsin 53706, USA}

\author{A.~Norambuena}
 
\affiliation{Centro de Investigaci\'on DAiTA Lab, Facultad de Estudios Interdisciplinarios, Universidad Mayor, Santiago, Chile}

\author{J.~R.~Maze}
 
\affiliation{Instituto de F\'isica, Pontificia Universidad Cat\'olica de Chile, Casilla 306, Santiago, Chile}

\affiliation{Centro de Investigaci\'on en Nanotecnolog\'ia y Materiales Avanzados, Pontificia Universidad Cat\'olica de Chile, Santiago, Chile}

\author{S.~Kolkowitz}
\email{kolkowitz@wisc.edu}
 
\affiliation{Department of Physics, University of Wisconsin, Madison, Wisconsin 53706, USA}

\maketitle

\section{Derivation of phonon-induced relaxation rates}

Here we derive expressions for the relaxation rates due to two-phonon processes consistent with the exponential and \(T^{5}\) terms in Eq.~6 of the main text. The calculation considers processes involving acoustic phonons in the bulk and quasilocalized phonons, which have together been shown to provide significant contributions to the experimentally observed temperature dependence of the lifetime of \(\ket{0}\) at \(T \gtrsim 50 \text{ K}\) \cite{Jarmola2012, Norambuena2018}. In addition, we demonstrate that due to time-reversal symmetry that two-acoustic-phonon Raman processes do not contribute to \(\gamma\) in the standard theory while two-quasilocalized-phonon Orbach-like processes can in general contribute to both the qutrit and qubit relaxation rates.

The calculation proceeds by application of Fermi's golden rule to second order, treating the interaction between the NV center and the diamond lattice as a perturbation. The eigenbasis is simultaneously diagonal in the Hamiltonian governing the NV and the Hamiltonian governing the lattice.

\subsection{NV Hamiltonian}

We consider an NV ground-state triplet Hamiltonian of the form 
\begin{equation}\label{spin_hamiltonian}
    H_{\text{NV}}/h = D_{\text{gs}}S_{z}^{2} + g\mu_{\text{B}}\boldsymbol{B} \cdot \boldsymbol{S},
\end{equation}
where \(h=2\pi\hbar\) is the Planck constant, \(D_{\text{gs}}=2.87 \text{ GHz}\) is the ground-state zero-field splitting, and \(g\mu_{\text{B}}=2.8 \text{ MHz/G}\) is the NV electronic spin gyromagnetic ratio. The spin-1 operators are denoted \(\boldsymbol{S}=(S_{x},S_{y},S_{z})\) and the magnetic field is \(\boldsymbol{B}\). 

\subsection{Lattice Hamiltonian}

The lattice Hamiltonian for a perfect lattice (i.e. a lattice without any defects) is
\begin{equation}\label{l_hamiltonian_position}
    H_{\text{l}} = \frac{1}{2m} \sum_{i\alpha} p_{i\alpha}^{2} +\frac{1}{2} \sum_{ii'\alpha\alpha'} \Phi(\boldsymbol{\Delta}_{ii'}) \, u_{i\alpha} \, u_{i'\alpha'},
\end{equation}
where \(i\) indexes the atoms comprising the lattice, and \(\alpha\) spans the Cartesian components \((x,y,z)\). The mass of a carbon atom, which comprises the lattice, is denoted \(m\). The atomic momenta and equilibrium displacements are \(p_{i\alpha}\) and \(u_{i\alpha}\). 
The interaction between atoms is described to second order in the displacements with isotropic coupling \(\Phi(\boldsymbol{\Delta}_{ii'})\), where \(\boldsymbol{\Delta}_{ii'} = \boldsymbol{R}_{i} - \boldsymbol{R}_{i'}\) is the distance between the equilibrium positions of the \(i\)th and \(i'\)th atoms. Though this Hamiltonian is insufficient on its own to describe quasilocalized modes, which arise due to the presence of defects in the lattice, the quasilocalized modes may be considered perturbations of existing modes within the bulk of the crystal since their frequencies fall in the band of bulk vibrational modes \cite{Gali2011, Zhang2011, Alkauskas2014}. Quasilocalized modes will be treated phenomenologically in a subsequent section. Examination of the perfect lattice Hamiltonian reveals some important properties of acoustic phonons and demonstrates the general method by which lattice vibrations can be quantized.

We diagonalize the lattice Hamiltonian with discrete Fourier transforms over the \(N\) atoms that comprise the lattice:
\begin{gather}
    \label{p_fourier}p_{i\alpha} = \frac{1}{\sqrt{N}} \sum_{k} p_{k} e_{k\alpha} \exp(i \boldsymbol{k} \cdot \boldsymbol{R}_{i}),\\
    \label{u_fourier}u_{i\alpha} = \frac{1}{\sqrt{N}} \sum_{k} u_{k\alpha} e_{k\alpha} \exp(i \boldsymbol{k} \cdot \boldsymbol{R}_{i}),
\end{gather}
where \(k\) indexes the vibrational mode with wavevector \(\boldsymbol{k}\) and polarization unit vector \(\boldsymbol{e}_{k}\). The \(\alpha\)th component of \(\boldsymbol{e}_{k}\) is denoted \(e_{k\alpha}\). We may then abbreviate the wavevector space momentum and position as
\begin{gather}
    p_{k\alpha} = p_{k} e_{k\alpha},\\
    u_{k\alpha} = u_{k} e_{k\alpha}.
\end{gather}

With \(-k\) indicating a vibrational mode with the same polarization but opposite wavevector as \(k\), the transformation of Eqs. \ref{p_fourier} and \ref{u_fourier} results in the Hamiltonian 
\begin{equation}\label{l_hamiltonian_wavevector}
    H_{\text{l}} = \sum_{k\alpha} \left( \frac{1}{2m} p_{k\alpha} \, p_{-k\alpha} +\frac{m \omega_{k}^{2}}{2} u_{k\alpha} u_{-k\alpha} \right),
\end{equation}
where \(\hbar \omega_{k}\) describes the energy associated with one quantum of the vibrational mode \(k\). We can rewrite the Hamiltonian using the creation and annihilation operators \(b^{\dagger}\) and \(b\), which obey
\begin{gather}
    \label{p_ladder}p_{k\alpha} = i\sqrt{\frac{\hbar m \omega_{k}}{2}} \left(b^{\dagger}_{-k} - b_{k}\right)e_{k\alpha},\\
    \label{u_ladder}u_{k\alpha} = \sqrt{\frac{\hbar}{2 m \omega_{k}}} \left(b^{\dagger}_{-k} + b_{k}\right)e_{k\alpha}.
\end{gather}
Using Eqs. \ref{p_ladder} and \ref{u_ladder}, we obtain
\begin{align}
    \label{l_hamiltonian_phonon} H_{\text{l}} &= \sum_{k}\hbar \omega_{k} \left(b^{\dagger}_{k}b_{k} +\tfrac{1}{2}\right),
\end{align}
which gives the energy of the lattice in terms of the occupation numbers of the vibrational modes. The eigenstates of the Hamiltonian are Fock, or number, states. A single quantum of a mode can be treated as a bosonic quasiparticle, the phonon. In general, the quantization of lattice vibrations described by Eq. \ref{l_hamiltonian_phonon} can be used as the Hamiltonian for a more complicated lattice by extending the range of \(k\) to cover vibrational modes beyond those of a perfect lattice.

\subsection{Interaction Hamiltonian}

A phenomenological model of the spin-lattice interaction to second order in the atomic displacements \(u_{i\alpha}\) is described by:
\begin{equation}\label{sl_hamiltonian_position}
    H_{\text{sl}} = \left(\sum_{mm' i \alpha}\lambda_{mm' i \alpha}\ket{m}\bra{m'}u_{i\alpha}\right) + \left(\sum_{mm' i i' \alpha \alpha'}\lambda_{mm' i i' \alpha \alpha'}\ket{m}\bra{m'}u_{i\alpha}u_{i'\alpha'}\right)
\end{equation}
where \(\ket{m}\) and \(\ket{m'}\) are eigenstates of the spin Hamiltonian. The \(\lambda_{mm' i \alpha}\) are coupling constants to first order in the atomic displacements and the \(\lambda_{mm' i i' \alpha \alpha'}\) are coupling constants to second order in the atomic displacements. To avoid an unnecessary proliferation of terms, we will consider only the first-order contributions as we demonstrate how this interaction Hamiltonian may be expressed in terms of the phonon creation and annihilation operators. 
Excluding the second-order terms, Eq. \ref{sl_hamiltonian_position} can also be written 
\begin{equation}\label{sl_hamiltonian_position_NV}
    H_{\text{sl}} = \sum_{\beta i \alpha}\lambda_{\beta i \alpha}F_{\beta}u_{i\alpha},
\end{equation}
where the \(F_{\beta}\) are linear combinations of the operators \(\ket{m}\bra{m'}\), \(\beta\) indexes the combinations, and the \(\lambda_{\beta i \alpha}\) are coupling constants. For the NV ground-state triplet, consideration of time-reversal and spatial symmetries imposes the restrictions
\begin{gather}
    \beta = \left(z,x',y',x,y\right),\\
    (F_{z},\,F_{x'},\,F_{y'},\,F_{x},\,F_{y}) = (S^{2}_{z},\,\{S_{x},S_{z}\},\,\{S_{y},S_{z}\},\,S^{2}_{y}-S^{2}_{x},\,\{S_{x},S_{y}\}),
\end{gather}
which may be demonstrated by either linear algebra or group theory \cite{Udvarhelyi2018}. 
Working from Eq. \ref{sl_hamiltonian_position_NV}, the Fourier transform given by Eq. \ref{u_fourier} yields
\begin{align}
    H_{\text{sl}} = \sum_{\beta i k \alpha}\frac{1}{\sqrt{N}} \lambda_{\beta i \alpha} \exp(i \boldsymbol{k} \cdot \boldsymbol{R}_{i}) F_{\beta} u_{k\alpha}.
\end{align}
Using Eq. \ref{u_ladder} to expand in terms of the creation and annihilation operators, 
\begin{align}
    H_{\text{sl}} = \sum_{\beta i k \alpha}\sqrt{\frac{\hbar}{2Nm\omega_{k}}} \lambda_{\beta i \alpha} \exp(i \boldsymbol{k} \cdot \boldsymbol{R}_{i}) F_{\beta} \left(b^{\dagger}_{-k} + b_{k}\right)e_{k\alpha}.
\end{align}
We define
\begin{align}
    \lambda_{\beta k} &= \sum_{i\alpha}\sqrt{\frac{\hbar}{2Nm\omega_{k}}} \lambda_{\beta i \alpha} e_{k\alpha} \exp(i \boldsymbol{k} \cdot \boldsymbol{R}_{i})
\end{align}
and obtain the spin-phonon Hamiltonian 
\begin{equation}\label{sl_hamiltonian_phonon}
    H_{\text{sl}} = \sum_{\beta k}\lambda_{\beta k}F_{\beta}\left(b^{\dagger}_{-k} + b_{k}\right).
\end{equation}
It can easily be shown that the spin-phonon Hamiltonian including second-order terms is analogously
\begin{equation}\label{sl_hamiltonian_phonon_bilinear}
    H_{\text{sl}} = \left(\sum_{\beta k}\lambda_{\beta k}F_{\beta}\left(b^{\dagger}_{-k} + b_{k}\right)\right) + \left(\sum_{\beta k k'}\lambda_{\beta k k'}F_{\beta}\left(b^{\dagger}_{-k} + b_{k}\right)\left(b^{\dagger}_{-k'} + b_{k'}\right)\right),
\end{equation}
where 
\begin{equation}
    \lambda_{\beta k k'} = \sum_{ii'\alpha\alpha'}\frac{\hbar}{2Nm\sqrt{\omega_{k}\omega_{k'}}} \lambda_{\beta i i' \alpha \alpha'} e_{k\alpha} \exp(i \boldsymbol{k} \cdot \boldsymbol{R}_{i}) e_{k'\alpha'} \exp(i \boldsymbol{k}' \cdot \boldsymbol{R}_{i'})
\end{equation}

We note that if we assume the most significant contributions to \(\lambda_{\beta k}\) come from atoms close to the defect, then for long wavelength phonons in the context of the spin-phonon Hamiltonian to first order in the atomic displacements we can write
\begin{equation}\label{long_wavelength}
    \lambda_{\beta k} \approx \sum_{i\alpha}\sqrt{\frac{\hbar}{2Nm\omega_{k}}} \lambda_{\beta i \alpha} e_{k\alpha}\left(1+i \boldsymbol{k} \cdot \boldsymbol{R}_{i}\right).
\end{equation}
Because the system is invariant under translation,
\begin{gather}
    \label{sum_lambdas} \sum_{i} \lambda_{\beta i \alpha} = 0,
\end{gather}
and Eq. \ref{long_wavelength} simplifies to
\begin{align}\label{long_simp}
    \lambda_{\beta k} &= \sum_{i\alpha}\sqrt{\frac{\hbar\omega_{k}}{2Nmv_{\text{s}}^{2}}} \lambda_{\beta i \alpha} e_{k\alpha} \left(i\hat{\boldsymbol{k}} \cdot \boldsymbol{R}_{i}\right),
\end{align}
where \(\hat{\boldsymbol{k}}\) is the unit vector along \(\boldsymbol{k}\), \(v_{\text{s}}\) is the speed of sound in diamond, and we have used the dispersion relation for acoustic phonons \(\omega_{k} = v_{\text{s}} \abs{\boldsymbol{k}}\). Eq. \ref{long_simp} displays the proportionality \(\lambda_{\beta k} \propto \sqrt{\omega_{k}}\), a property which will be invoked to obtain the \(T^{5}\) scaling of two-acoustic-phonon Raman processes. 

\subsection{Evaluation of Fermi's golden rule for Raman processes}

From time-dependent perturbation theory, Fermi's golden rule allows us to calculate \(\Gamma_{i \rightarrow f}\), the transition rate from an initial state \(\ket{i}\) to a final continuum state \(\ket{f}\). To second order,
\begin{align}\label{fermi}
    \Gamma_{i \rightarrow f} = \frac{2\pi}{\hbar} \abs{V_{fi}+\sum_{m}\frac{V_{fm}V_{mi}}{E_{i}-E_{m}}}^{2}\delta(E_{i}-E_{f}),
\end{align}
where \(V_{\alpha\beta}=\bra{\alpha}V\ket{\beta}\) are the matrix elements of the perturbation, \(E_{\alpha}\) is the energy of \(\ket{\alpha}\), and \(\ket{m}\) is an intermediate state. 

Treating the spin-phonon Hamiltonian as a perturbation, the eigenstates of the system are described by the tensor product of the NV Hamiltonian eigenstates and the lattice Fock states. 
We assume the initial Fock state occupation numbers \(n_{k}\) are described by the Bose-Einstein distribution
\begin{equation}
    n_{k} = \frac{1}{\exp(\hbar\omega_{k}/k_{\text{B}}T)-1}
\end{equation}
for temperature \(T\), where \(k_{\text{B}}\) is the Boltzmann constant. We consider all final states with the NV component of interest, regardless of the vibrational mode occupation numbers. That is, the phonon-induced NV transition rate is given by
\begin{equation}\label{Gamma_sum}
    \Gamma_{m_{s} \leftrightarrow m_{s}'} = \sum_{n_{k}',n_{k'}',...} \Gamma_{m_{s},n_{k},n_{k'},...}^{m_{s}',n_{k}',n_{k'}',...}
\end{equation}
where \(\Gamma_{m_{s},n_{k},n_{k'},...}^{m_{s}',n_{k}',n_{k'}',...}\) is the transition between an initial state \(\ket{m_{s}} \otimes \ket{...,n_{k},...,n_{k'},...}\) to a final state \(\ket{m_{s}'} \otimes \ket{...,n_{k}',...,n_{k'}',...}\). 

Because the majority of phonons at room temperature have energies much greater than the splitting between the levels of the NV ground-state triplet and the NV has no low-lying excited states, relatively few phonons satisfy the energy requirements for direct two-phonon processes. Therefore, we consider only Raman processes, in which one phonon is absorbed and another is emitted. 
The NV transition rate due solely to Raman processes is
\begin{equation}\label{Gamma_sum_simp}
    \Gamma_{m_{s} \leftrightarrow m_{s}'} = \sum_{kk'} \Gamma_{m_{s},n_{k},n_{k'}}^{m_{s}',n_{k}-1,n_{k'}+1}.
\end{equation}
Evaluating Eq. \ref{Gamma_sum_simp} using the spin-phonon Hamiltonian of Eq. \ref{sl_hamiltonian_phonon_bilinear}, we obtain a general expression for the transition rate associated with two-phonon Raman processes with intermediate states restricted to the multiplet containing the initial and final states:
\begin{align}\label{raman_rate_long_bilinear}
    \Gamma_{m_{s} \leftrightarrow m_{s}'} &= \frac{2\pi}{\hbar} \sum_{kk'} n_{k}\left(n_{k'}+1\right)\abs{H_{m_{s}m_{s}'kk'} + \sum_{m_{s}''}\left( \frac{H_{m_{s}'m_{s}''k'}H_{m_{s}''m_{s}k}}{\Delta_{m_{s}m_{s}''}+\hbar\omega_{k}} + \frac{H_{m_{s}'m_{s}''k}H_{m_{s}''m_{s}k'}}{\Delta_{m_{s}m_{s}''}-\hbar\omega_{k'}} \right)}^{2} \, \delta(\Delta_{m_{s}m_{s}'} + \hbar\omega_{k} - \hbar\omega_{k'})
\end{align}
where the matrix element \(H_{m_{s}m_{s}'kk'}\)  and \(H_{m_{s}m_{s}'k}\) are defined
\begin{gather}
    H_{m_{s}m_{s}'kk'} = \bra{m_{s}}\sum_{\beta}\lambda_{\beta k k'}F_{\beta}\ket{m_{s}'},\\
    H_{m_{s}m_{s}'k} = \bra{m_{s}}\sum_{\beta}\lambda_{\beta k}F_{\beta}\ket{m_{s}'},
\end{gather}
and
\begin{equation}
    \Delta_{m_{s}m_{s}'} = \bra{m_{s}}H_{\text{NV}}\ket{m_{s}}-\bra{m_{s}'}H_{\text{NV}}\ket{m_{s}'}
\end{equation}
is the energy difference between the NV states \(\ket{m_{s}}\) and \(\ket{m_{s}'}\). Evaluating the squared magnitude in Eq.~\ref{raman_rate_long_bilinear},
\begin{align}\label{raman_rate_long_bilinear_expanded}
    \nonumber\Gamma_{m_{s} \leftrightarrow m_{s}'} &= \frac{2\pi}{\hbar} \sum_{kk'} n_{k}\left(n_{k'}+1\right) \, \delta(\Delta_{m_{s}m_{s}'} + \hbar\omega_{k} - \hbar\omega_{k'})\\
    \nonumber&\quad \times \left( \abs{H_{m_{s}m_{s}'kk'}}^{2} + 2 \Re{H_{m_{s}m_{s}'kk'}^{*}\sum_{m_{s}''}\left( \frac{H_{m_{s}'m_{s}''k'}H_{m_{s}''m_{s}k}}{\Delta_{m_{s}m_{s}''}+\hbar\omega_{k}} + \frac{H_{m_{s}'m_{s}''k}H_{m_{s}''m_{s}k'}}{\Delta_{m_{s}m_{s}''}-\hbar\omega_{k'}} \right)}
    \vphantom{\abs{\sum_{m_{s}''}\left( \frac{H_{m_{s}'m_{s}''k'}H_{m_{s}''m_{s}k}}{\Delta_{m_{s}m_{s}''}+\hbar\omega_{k}} + \frac{H_{m_{s}'m_{s}''k}H_{m_{s}''m_{s}k'}}{\Delta_{m_{s}m_{s}''}-\hbar\omega_{k'}} \right)}^{2}}\right. 
    \\
    &\qquad\quad \left. + \abs{\sum_{m_{s}''}\left( \frac{H_{m_{s}'m_{s}''k'}H_{m_{s}''m_{s}k}}{\Delta_{m_{s}m_{s}''}+\hbar\omega_{k}} + \frac{H_{m_{s}'m_{s}''k}H_{m_{s}''m_{s}k'}}{\Delta_{m_{s}m_{s}''}-\hbar\omega_{k'}} \right)}^{2}\right).
\end{align}
We see that two-phonon Raman processes arise in three ways according to the three major terms in the large parenthesis in Eq.~\ref{raman_rate_long_bilinear}. The first term,
\begin{equation}\nonumber
    \abs{H_{m_{s}m_{s}'kk'}}^{2},
\end{equation}
corresponds to the second-order terms in the spin-phonon Hamiltonian taken to first order in perturbation theory. The mixed second term,
\begin{equation}\nonumber
    2\Re{H_{m_{s}m_{s}'kk'}^{*}\sum_{m_{s}''}\left( \frac{H_{m_{s}'m_{s}''k'}H_{m_{s}''m_{s}k}}{\Delta_{m_{s}m_{s}''}+\hbar\omega_{k}} + \frac{H_{m_{s}'m_{s}''k}H_{m_{s}''m_{s}k'}}{\Delta_{m_{s}m_{s}''}-\hbar\omega_{k'}} \right)},
\end{equation}
results from the interference of the first- and second-order terms in the spin-phonon Hamiltonian. The third term,
\begin{equation}\nonumber
    \abs{\sum_{m_{s}''}\left( \frac{H_{m_{s}'m_{s}''k'}H_{m_{s}''m_{s}k}}{\Delta_{m_{s}m_{s}''}+\hbar\omega_{k}} + \frac{H_{m_{s}'m_{s}''k}H_{m_{s}''m_{s}k'}}{\Delta_{m_{s}m_{s}''}-\hbar\omega_{k'}} \right)}^{2}
\end{equation}
corresponds to the first-order terms in the spin-phonon Hamiltonian taken to second order in perturbation theory.

For the remainder of this supplement, we explicitly consider only the third term, since this is the premise of both the exponential and \(T^{5}\) terms in Eq. 6 of the main text. We will comment on the possible contributions of the other terms in a subsequent section. With this simplification, we can write
\begin{align}\label{raman_rate_long}
    \Gamma_{m_{s} \leftrightarrow m_{s}'} &= \frac{2\pi}{\hbar} \sum_{kk'} n_{k}\left(n_{k'}+1\right)\abs{\sum_{m_{s}''}\left( \frac{H_{m_{s}'m_{s}''k'}H_{m_{s}''m_{s}k}}{\Delta_{m_{s}m_{s}''}+\hbar\omega_{k}} + \frac{H_{m_{s}'m_{s}''k}H_{m_{s}''m_{s}k'}}{\Delta_{m_{s}m_{s}''}-\hbar\omega_{k'}} \right)}^{2} \, \delta(\Delta_{m_{s}m_{s}'} + \hbar\omega_{k} - \hbar\omega_{k'}).
\end{align}
This generic second-order two-phonon Raman process may be interpreted as follows. 
For each intermediate NV state \(m_{s}''\), the terms in the sum in Eq. \ref{raman_rate_long} represent an absorption-followed-by-emission variant (first term in the sum, depicted in Fig.~4(a) of the main text) and an emission-followed-by-absorption variant (second term in the sum, depicted in Fig.~4(b) of the main text) of the two-phonon process. As we will see, it is important that these processes are not identical, but are associated with different denominators and may interfere with one another. Eq. \ref{raman_rate_long} will be the starting point for our analyses of transitions involving acoustic phonons and quasilocalized phonons, which will be treated separately. 

\subsection{Properties of the generic second-order two-phonon Raman process under time-reversal}

Before considering the effects of different types of phonons, it is helpful to examine the properties of the generic two-phonon process under time-reversal. The restriction that the spin-lattice Hamiltonian is symmetric under time-reversal excludes odd powers of the spin operators from appearing in the Hamiltonian. The operator \(\sum_{\beta}\lambda_{\beta k} F_{\beta}\) is thus also symmetric under time reversal. With \(\Theta\) denoting the time-reversal operator, the \(S_{z}\) eigenstates obey
\begin{equation}
    \Theta \ket{m_{s}} = (-1)^{m_{s}} \ket{-m_{s}}.
\end{equation}
Thus the matrix elements \(H_{m_{s}m_{s}'k}\) follow
\begin{align}
    \bra{m_{s}} \sum_{\beta}\lambda_{\beta k} F_{\beta} \ket{m_{s}'} &= (-1)^{m_{s}+m_{s}'} \bra{-m_{s}} \sum_{\beta}\lambda_{\beta k} F_{\beta} \ket{-m_{s}'}^{*}\\
    &= (-1)^{m_{s}+m_{s}'} \bra{-m_{s}'} \sum_{\beta}\lambda_{\beta k} F_{\beta} \ket{-m_{s}},
\end{align}
where we have used the antiunitarity of time-reversal and taken the Hermitian conjugate.
More concisely,
\begin{equation}\label{t_reversal_mat_els}
    H_{m_{s}m_{s}'k} = (-1)^{m_{s}+m_{s}'} H_{-m_{s}'-m_{s}k}.
\end{equation}

Inspecting Eq.~\ref{raman_rate_long}, it may seem appropriate to make the approximation 
\begin{equation}\label{walker_approx}
    \abs{\sum_{m_{s}''}\left( \frac{H_{m_{s}'m_{s}''k'}H_{m_{s}''m_{s}k}}{\Delta_{m_{s}m_{s}''}+\hbar\omega_{k}} + \frac{H_{m_{s}'m_{s}''k}H_{m_{s}''m_{s}k'}}{\Delta_{m_{s}m_{s}''}-\hbar\omega_{k'}} \right)}^{2} \approx \frac{1}{(\hbar\omega_{k})^{2}} \abs{\sum_{m_{s}''}\left( H_{m_{s}'m_{s}''k'}H_{m_{s}''m_{s}k} - H_{m_{s}'m_{s}''k}H_{m_{s}''m_{s}k'} \right)}^{2}
\end{equation}
where we have assumed the delta function and used the fact that the energy scale of the NV is small in comparison to the energies of typical phonons at room temperature. In this approximation, \(\Gamma_{m_{s} \leftrightarrow m_{s}'}\) is only nonzero if the quantity
\begin{equation}
    A_{m_{s}m_{s}'kk'} = \sum_{m_{s}''}\left( H_{m_{s}'m_{s}''k'}H_{m_{s}''m_{s}k} - H_{m_{s}'m_{s}''k}H_{m_{s}''m_{s}k'}\right)
\end{equation}
is nonzero in general. We will now show that \(A_{m_{s}m_{s}'kk'}\) is exactly zero for the transition \(\ket{m_{s}} \leftrightarrow \ket{-m_{s}}\) by using the time-reversal symmetry described by Eq.~\ref{t_reversal_mat_els}. 
The demonstration will in fact apply broadly to \(\ket{m_{s}} \leftrightarrow \ket{-m_{s}}\) transitions within a multiplet of an integer-spin system. The terms of the sum over \(m_{s}''\) in \(A_{m_{s}-m_{s}kk'}\) may be divided into two simple cases.

Case 1, \(m_{s}'' \neq 0\): If \(m_{s}'' \neq 0\), then  
\begin{align}
    H_{-m_{s}m_{s}''k'}H_{m_{s}''m_{s}k} &= (-1)^{-m_{s}+m_{s}''+m_{s}''+m_{s}}H_{-m_{s}''m_{s}k'}H_{-m_{s}-m_{s}''k}\\
    \label{t_sym_mat_els} &= H_{-m_{s}-m_{s}''k}H_{-m_{s}''m_{s}k'}.
\end{align}
The opposite of this term enters \(A_{m_{s}-m_{s}kk'}\) via the intermediate \(-m_{s}''\). Therefore, for each term associated with a particular \(m_{s}''\), there is a cancellation that occurs with a term from \(-m_{s}''\) and the total contribution from intermediates \(m_{s}'' \neq 0\) is 0.

Case 2, \(m_{s}'' = 0\): This is just a special case of Eq.~\ref{t_sym_mat_els}. We have 
\begin{align}
    H_{-m_{s}0k}H_{0m_{s}k'} &= H_{-m_{s}0k'}H_{0m_{s}k}
\end{align}
for which the opposite term comes from the remaining part of the \(m_{s}'' = 0\) contribution. 

Because \(A_{m_{s}-m_{s}kk'}=0\), it is clear that the approximation of Eq.~\ref{walker_approx} is insufficient to explain relaxation on \(\ket{m_{s}} \leftrightarrow \ket{-m_{s}}\) transitions. More specifically, nonzero relaxation rates for a two-phonon process on \(\ket{m_{s}} \leftrightarrow \ket{-m_{s}}\) transitions are contingent on the difference between the denominators for the absorption-followed-by-emission and emission-followed-by-absorption variants of the process. 
The finding \(A_{m_{s}-m_{s}kk'}=0\) can also be easily verified for the specific case of the NV's qutrit transition by evaluating the matrix elements explicitly. It should be noted that the same symmetry argument clearly does not hold if \(m_{s} \neq -m_{s}'\), as in the case of the NV's qubit transition. A similar symmetry argument to the one presented here has historically been applied to half-integer-spin systems, where it is referred to as Van Vleck cancellation \cite{Van1940}. 

\subsection{Two-acoustic-phonon Raman processes}

The two-acoustic-phonon Raman process with \(T^{5}\) thermal scaling presented by Walker in \cite{Walker1968} depends on the approximation of Eq.~\ref{walker_approx}, which we have just shown to be insufficient to describe relaxation on the qutrit transition. For completeness, we will demonstrate the rest of the derivation here, showing how the temperature dependence of the process arises. We restrict consideration to acoustic vibrational modes (denoted aco) and make the approximation of Eq.~\ref{walker_approx}.
This yields
\begin{align}\label{aco_rate_supp}
    \Gamma_{m_{s} \leftrightarrow m_{s}'}^{\text{(aco)}} &= \frac{2\pi}{\hbar^{3}} \sum_{kk' \, \in \, \text{aco}} \frac{n_{k}\left(n_{k}+1\right)}{\omega_{k}^{2}}\abs{\sum_{m_{s}''}\left(H_{m_{s}'m_{s}''k'}H_{m_{s}''m_{s}k} - H_{m_{s}'m_{s}''k}H_{m_{s}''m_{s}k'}\right)}^{2} \, \delta(\hbar\omega_{k} - \hbar\omega_{k'}).
\end{align}

From here, our derivation will diverge slightly from the original derivation in \cite{Walker1968}. In particular, while the original derivation was made with arguments specific to a simple cubic lattice, we will generalize those arguments for an arbitrary lattice. Transforming from wavevector space back to position space and taking the long wavelength approximation of Eq. \ref{long_simp} allows the matrix elements \(H_{m_{s}m_{s}'k}\) to be expressed
\begin{align}\label{walker_mat_el}
    H_{m_{s}m_{s}'k} &\approx \sum_{i \alpha} \sqrt{\frac{\hbar\omega_{k}}{2Nmv_{\text{s}}^{2}}} e_{k\alpha} \left(i\hat{\boldsymbol{k}} \cdot \boldsymbol{R}_{i}\right) H_{m_{s}m_{s}'i \alpha}
\end{align}
where \(H_{m_{s}m_{s}'i \alpha}\) is the position space analog of \(H_{m_{s}m_{s}'k}\) defined
\begin{equation}
    H_{m_{s}m_{s}'i \alpha} = \bra{m_{s}}\sum_{\beta}\lambda_{\beta i \alpha}F_{\beta}\ket{m_{s}'}.
\end{equation}
Substitution of Eq. \ref{walker_mat_el} into Eq. \ref{aco_rate_supp} yields
\begin{equation}
    \Gamma_{m_{s} \leftrightarrow m_{s}'}^{\text{(aco)}} = \frac{\pi}{2\hbar^{2}N^{2}m^{2}v_{\text{s}}^{4}} \sum_{kk' \, \in \, \text{aco}} n_{k}\left(n_{k}+1\right) \, B_{m_{s}m_{s}'kk'} \, \delta(\omega_{k} - \omega_{k'})
\end{equation}
with 
\begin{equation}
    B_{m_{s}m_{s}'kk'} = \abs{\sum_{ii'\alpha\alpha'm_{s}''}e_{k\alpha} e_{k'\alpha'} \left(\hat{\boldsymbol{k}} \cdot \boldsymbol{R}_{i}\right) \left(\hat{\boldsymbol{k}}' \cdot \boldsymbol{R}_{i'}\right) \left(H_{m_{s}'m_{s}''i \alpha} H_{m_{s}''m_{s}i' \alpha'} - H_{m_{s}'m_{s}''i' \alpha'} H_{m_{s}''m_{s}i \alpha} \right)}^{2}.
\end{equation}
We observe that \(B_{m_{s}m_{s}'kk'}\) depends only on the spatial characteristics of the vibrational modes \(k\) and \(k'\). In the continuum limit, we may therefore reasonably replace \(B_{m_{s}m_{s}'kk'}\) by a spatial average \(B_{m_{s}m_{s}'}\) defined
\begin{multline}
    B_{m_{s}m_{s}'} = \frac{1}{144\pi^{2}} \sum_{\gamma \gamma'} \iint \bigg|\sum_{ii'\alpha\alpha'm_{s}''} \left(\boldsymbol{P}_{\gamma}(\boldsymbol{n}) \cdot \boldsymbol{e}_{\alpha}\right) \left(\boldsymbol{P}_{\gamma'}(\boldsymbol{n}') \cdot \boldsymbol{e}_{\alpha}\right) \left(\boldsymbol{n} \cdot \boldsymbol{R}_{i}\right) \left(\boldsymbol{n}' \cdot \boldsymbol{R}_{i'}\right) \\
    \left(H_{m_{s}'m_{s}''i \alpha} H_{m_{s}''m_{s}i' \alpha'} - H_{m_{s}'m_{s}''i' \alpha'} H_{m_{s}''m_{s}i \alpha}\right)\bigg|^{2} \, dS \, dS'
\end{multline}
where the integrals each cover the unit sphere and \(\gamma\) and \(\gamma'\) each index the one longitudinal and two transverse polarizations. The unit vector \(\boldsymbol{n}\) is the normal associated with the surface element \(dS\), and the function \(\boldsymbol{P}_{\gamma}(\boldsymbol{n})\) returns the \(\gamma\)th polarization unit vector associated with \(\boldsymbol{n}\).

We take the continuum limit with the transformations 
\begin{gather}
    \omega_{k} \rightarrow \omega,\\
    n_{k} \rightarrow n(\omega) = \frac{1}{\exp(\hbar \omega / k_{\text{B}}T)-1},\\
    B_{m_{s}m_{s}'kk'} \rightarrow B_{m_{s}m_{s}'}.
\end{gather}
The sums over vibrational modes become integrals over frequencies with measure provided by the density of states of acoustic phonons. We assume the Debye model density of states given by
\begin{equation}\label{dos2}
    D(\omega) = \frac{3V}{2\pi^{2}v_{\text{s}}^{3}}\omega^{2} \, \Pi\left(\frac{\omega}{\omega_{\text{D}}}\right),
\end{equation}
where \(\omega_{\text{D}}\) is the Debye frequency and \(V\) is the volume of the crystal. We have defined the rectangle function
\begin{equation}
    \Pi(x) = 
    \begin{cases}
        1, & 0<x<1\\
        0, & \text{otherwise}.
    \end{cases}
\end{equation}
We obtain
\begin{equation}
    \Gamma_{m_{s} \leftrightarrow m_{s}'}^{\text{(aco)}} = \frac{9V^{2}}{8 \pi^{3} \hbar^{2}N^{2}m^{2}v_{\text{s}}^{10}} \, B_{m_{s}m_{s}'} \iint n(\omega)\left(n(\omega)+1\right)\, \omega^{4} \, \delta(\omega - \omega') \, \Pi\left(\frac{\omega}{\omega_{\text{D}}}\right) \, d\omega \, d\omega'.
\end{equation}
Evaluating the integral over \(\omega'\),
\begin{equation}\label{T5_pre_var_change}
    \Gamma_{m_{s} \leftrightarrow m_{s}'}^{\text{(aco)}} = \frac{9}{8 \pi^{3} \hbar^{2}d^{2}m^{2}v_{\text{s}}^{10}} \, B_{m_{s}m_{s}'} \int_{0}^{\omega_{\text{D}}} n(\omega)\left(n(\omega)+1\right) \omega^{4} \, d\omega
\end{equation}
where \(d=N/V\) is the atomic number density.

With the change of variable
\begin{gather}
    x = \frac{\hbar \omega}{k_{\text{B}}T},\\
    x_{\text{D}} = \frac{\hbar \omega_{\text{D}}}{k_{\text{B}}T},\\
    \eta(x) = \frac{1}{\exp(x)-1},
\end{gather}
Eq. \ref{T5_pre_var_change} can be rewritten
\begin{equation}\label{T5_post_var_change}
    \Gamma_{m_{s} \leftrightarrow m_{s}'}^{\text{(aco)}} = \frac{9k_{\text{B}}^{5}T^{5}}{8 \pi^{3} \hbar^{7}n^{2}m^{2}v_{\text{s}}^{10}} \, B_{m_{s}m_{s}'} \int_{0}^{x_{\text{D}}} \eta(x)\left(\eta(x)+1\right) x^{4} \, dx.
\end{equation}
Because the integral depends only weakly on temperature for \(T \lesssim 300 \text{ K}\), Eq. \ref{T5_post_var_change} exhibits a thermal scaling of approximately \(T^{5}\) at room temperature. In the limit \(\omega_{\text{D}} \rightarrow \infty\), Eq. \ref{T5_post_var_change} exhibits an exact \(T^{5}\) scaling. 

We now enumerate three different schemes by which acoustic phonons may contribute to two-phonon relaxation as alternatives to the derivation shown in this section.

Alternative 1, restrict intermediate NV states to higher-lying states beyond the ground-state triplet: Because the splittings associated with these states are much greater than typical phonon energies at room temperature, we approximate \(\Delta_{m_{s}m_{s}''} \gg \hbar \omega_{k}\) for any \(m_{s}\), \(m_{s}''\), and \(\omega_{k}\). This results in a scaling of \(T^{7}\). 

Alternative 2, neglect the spatial characteristics of the acoustic phonon coupling constants \(\lambda_{\beta k}\): If we do not make the approximation of Eq.~\ref{walker_approx} and instead neglect the spatial characteristics of the coupling constants \(\lambda_{\beta k}\) (which is necessary to make the calculation tractable in this case), then \(H_{m_{s}'m_{s}''k'}H_{m_{s}''m_{s}k} = H_{m_{s}'m_{s}''k}H_{m_{s}''m_{s}k'}\) for phonons of the same energy. Now for any \(m_{s}\) and \(m_{s}'\) the rate \(\Gamma_{m_{s} \leftrightarrow m_{s}'}\) is only nonzero by the inequality of the denominators for the absorption-followed-by-emission and emission-followed-by-absorption variants in the generic two-phonon process of Eq. \ref{raman_rate_long}. This assumption results in a scaling of \(T^{3}\). 

Alternative 3, consider the spin-lattice Hamiltonian up to second order in the atomic displacements (see Eq.~\ref{raman_rate_long_bilinear_expanded}), which was originally proposed to explain quadrupolar nuclear spin-lattice relaxation \cite{Van1954}. The second-order contributions to the spin-phonon Hamiltonian taken to first order in perturbation theory exhibit a \(T^{7}\) thermal scaling for acoustic phonons, which has been shown to contribute negligibly to \(\Omega\) at room temperature \cite{Norambuena2018}. The time-reversal symmetry argument developed in the previous section does not apply to this mechanism since there is no interference of terms by which cancellation can occur. We leave exploration of the mixed term in Eq.~\ref{raman_rate_long_bilinear_expanded} for future work.

\subsection{Two-quasilocalized-phonon Orbach-like processes}

Quasilocalized vibrational modes are expected to couple more strongly to the NV spin than delocalized bulk modes because the atoms with the most significant displacements for a quasilocalized mode are in close proximity to the defect. In Ref. \cite{Alkauskas2014}, Alkausas \textit{et al.} demonstrated from first principles the existence of a vibrational resonance induced by the NV at \(\hbar\omega_{\text{loc}} = 65 \text{ meV}\) with a full width at half maximum (FWHM) of \(\hbar\Delta_{\text{loc}} = 32 \text{ meV}\), consistent with experimental measurements of NV photoluminescence and prior calculations of quasilocalized vibrational modes \cite{Zhang2011, Gali2011}. The effect of the quasilocalized modes comprising the resonance was phenomenologically modeled in Ref.~\cite{Norambuena2018} by including a delta function in the density of states. Because the resonance is a relatively wide feature, a less simplified model should account for its width. We demonstrate a model following this consideration here, and show that its temperature dependence is consistent with the exponential term in the empirical model of the NV qubit relaxation rate described by Eq.~6 of the main text. Because the quasilocalized modes comprising the resonance are associated with a single spatial vibrational pattern, we model the coupling of the NV spin to the quasilocalized modes by a Lorentzian which depends only on the mode frequency:
\begin{equation}
    \lambda_{\beta,\text{loc}}(\omega) = \lambda_{\beta,\text{loc}} \, S(\omega),
\end{equation}
where the \(\lambda_{\beta,\text{loc}}\) are constants describing the spatial character of the quasilocalized vibrational pattern and \(S(\omega)\) is the Lorentzian
\begin{equation}
    S(\omega) = \frac{1}{\pi}\frac{\tfrac{1}{2}\Delta_{\text{loc}}}{\left(\omega - \omega_{\text{loc}}\right)^{2} + \left(\tfrac{1}{2}\Delta_{\text{loc}}\right)^{2}}.
\end{equation}
The coupling constants are associated with the matrix elements
\begin{equation}
    H_{m_{s}m_{s}',\text{loc}} = \bra{m_{s}}\sum_{\beta}\lambda_{\beta,\text{loc}}F_{\beta}\ket{m_{s}'}.
\end{equation}
Evaluating Eq. \ref{raman_rate_long} in the continuum limit, the transition rate for Raman processes with quasilocalized phonons is therefore
\begin{align}
    \nonumber \Gamma_{m_{s} \leftrightarrow m_{s}'}^{\text{(loc)}} &= \frac{2\pi}{\hbar} \iint n(\omega)\left(n(\omega')+1\right)S(\omega)S(\omega')\abs{\sum_{m_{s}''}\left(\frac{1}{\Delta_{m_{s}m_{s}''}+\hbar\omega} + \frac{1}{\Delta_{m_{s}m_{s}''}-\hbar\omega'} \right)H_{m_{s}'m_{s}'',\text{loc}}H_{m_{s}''m_{s},\text{loc}}}^{2}\\
    \label{quasi_intermediate} &\quad \times  \, \delta(\Delta_{m_{s}m_{s}'} + \hbar\omega - \hbar\omega')D_{\text{loc}}(\omega)D_{\text{loc}}(\omega') \, d\omega \, d\omega'
\end{align}
where \(D_{\text{loc}}(\omega)\) is the density of states for quasilocalized modes, which can be assumed to take the form \(D_{\text{loc}}(\omega) = D_{\text{loc}} \, \omega^{2}\) as the quasilocalized mode frequencies are within the band of bulk modes.
The delta function in Eq. \ref{quasi_intermediate} fixes the value of \(\omega'\) to \(\omega'=\omega+\Delta_{m_{s}m_{s}'}/\hbar\). Because the transition frequencies associated with the NV are small in comparison to both the vibrational resonance FWHM and center frequency, we can take \(n(\omega') \approx n(\omega)\), \(S(\omega') \approx S(\omega)\), and \(D_{\text{loc}}(\omega') \approx D_{\text{loc}}(\omega)\). Performing the integral over \(\omega'\) then yields
\begin{align}\label{quasi_intermediate2}
    \Gamma_{m_{s} \leftrightarrow m_{s}'}^{\text{(loc)}} &\approx \frac{2\pi D_{\text{loc}}^{2}}{\hbar^{6}} \, \abs{\sum_{m_{s}''}\left(\Delta_{m_{s}m_{s}''}+\Delta_{m_{s}'m_{s}''} \right)H_{m_{s}'m_{s}'',\text{loc}}H_{m_{s}''m_{s},\text{loc}}}^{2} \int_{0}^{\omega_{\text{D}}} n(\omega)\left(n(\omega)+1\right)\left(S(\omega)\right)^{2} \, d\omega.
\end{align}
To demonstrate a simple temperature dependence for Eq. \ref{quasi_intermediate2}, we approximate \(n(\omega) \approx n(\omega_{\text{loc}})\) for phonon frequencies with major contributions to the integral and obtain
\begin{align}\label{quasi_simple_temp}
    \Gamma_{m_{s} \leftrightarrow m_{s}'}^{\text{(loc)}} &\approx \frac{2\pi D_{\text{loc}}^{2}}{\hbar^{6}} \, n(\omega_{\text{loc}})\left(n(\omega_{\text{loc}})+1\right) \, \abs{\sum_{m_{s}''}\left(\Delta_{m_{s}m_{s}''}+\Delta_{m_{s}'m_{s}''} \right)H_{m_{s}'m_{s}'',\text{loc}}H_{m_{s}''m_{s},\text{loc}}}^{2} \int_{0}^{\omega_{\text{D}}} \left(S(\omega)\right)^{2} \, d\omega.
\end{align}

The temperature dependence of \(\Gamma_{m_{s} \leftrightarrow m_{s}'}^{\text{(loc)}}\) can therefore be described by the Bose-Einstein distribution term \(n(\omega_{\text{loc}})\left(n(\omega_{\text{loc}})+1\right)\). For \(\hbar\omega_{\text{loc}} \gg k_{\text{B}}T\), we have \(n(\omega_{\text{loc}})\left(n(\omega_{\text{loc}})+1\right) \approx n(\omega_{\text{loc}})\), which is the temperature dependence typically invoked in studies of the lifetime of \(\ket{0}\) as a function of temperature \cite{Redman1991,Jarmola2012,Norambuena2018}. Treating \(\omega_{\text{loc}}\) as a fit parameter (i.e. equating \(\omega_{\text{loc}}\) with \(\Delta\) in Eq.~6 of the main text), Ref.~\cite{Jarmola2012} finds \(\omega_{\text{loc}}=73(4) \text{ meV}\), well within the line of the vibrational resonance at 65 meV.

Because the Orbach process exhibits the same temperature dependence as the process shown in this section, the exponential term in Eq.~6 of the main text has previously been identified explicitly as an Orbach process \cite{Redman1991} or referred to as the result of an Orbach-type process \cite{Jarmola2012,Norambuena2018}. It is important to recognize, however, that while the Raman process for quasilocalized phonons and the Orbach process have the same temperature scaling, the two represent physically different mechanisms. Specifically, the Orbach process is a two-phonon process in which the intermediate state of the defect is a low-lying excited state and the intermediate state of the combined defect-lattice system is real (i.e., its energy is the same as that of the initial and final states) \cite{Orbach1961, Finn1961}. In contrast the intermediate system state in the Raman process is virtual (i.e., its energy differs from that of the initial and final states). For these reasons we refer to the two-quasilocalized-phonon process discussed in this section as Orbach-like.

\begin{figure}[h]
\includegraphics[width=0.48\textwidth]{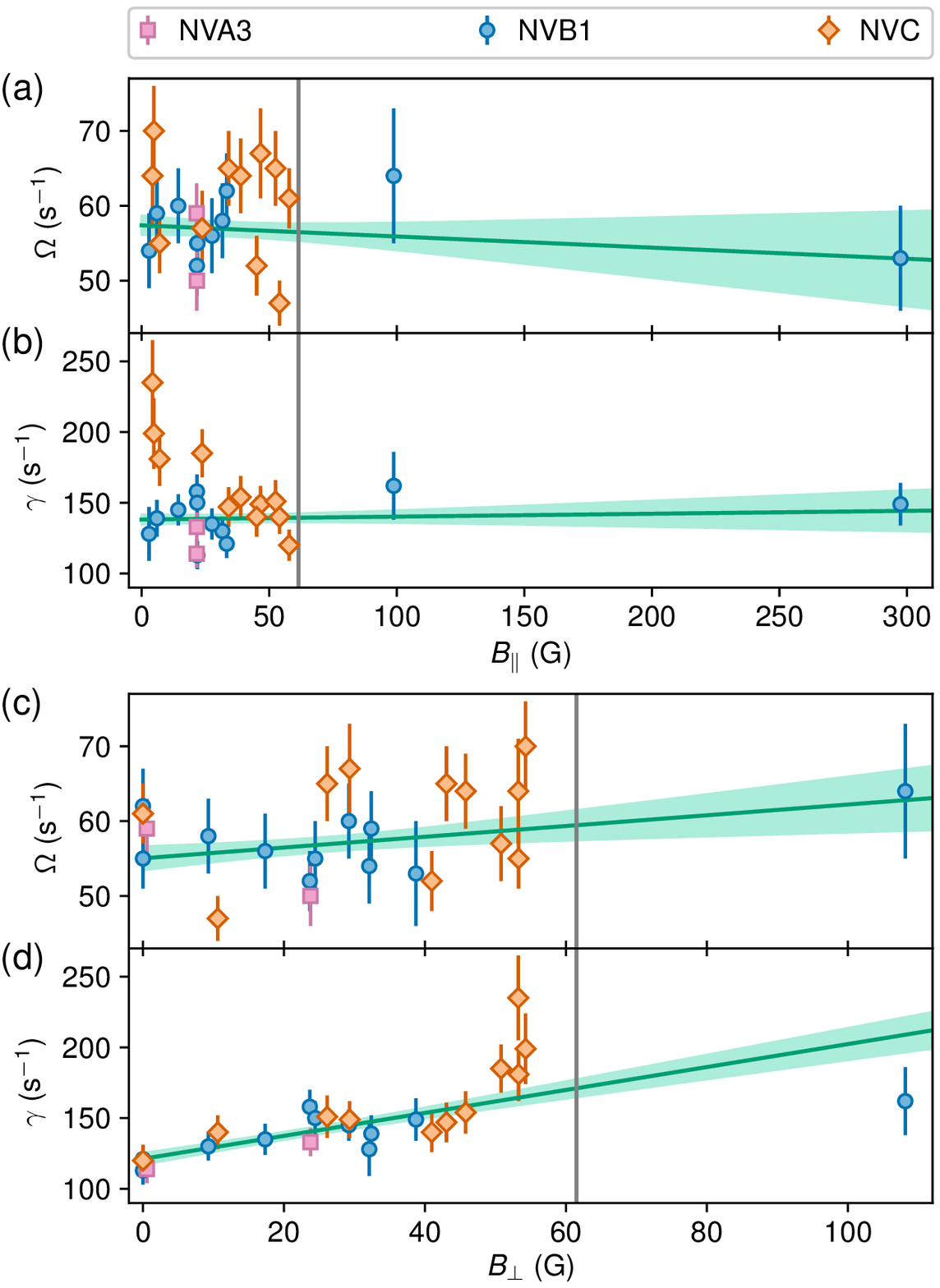}
\caption{\label{fig:rates_supp}Extended version of Fig.~3 of the main text, showing all data taken in this work for which the magnitude and orientation of the applied magnetic field is known. The vertical gray line at 65 G indicates the range of the data displayed in the main text version of this figure.
} 
\end{figure}

\begin{table*}[b]
\caption{\label{tab:complete_data}Complete set of data collected for this work. The magnitude and orientation of the applied magnetic field were determined with spin echo measurements. Information on the applied magnetic field is not included for measurements where spin echo was not performed.
}

\setlength{\tabcolsep}{9pt}
\begin{tabular}{cccccccc}
NV & \(\Delta_{\pm}\) (MHz) & \(B\) (G) & \(B_{\parallel}\) (G) & \(B_{\perp}\) (G) & \(\gamma\) (\(\times 10^{3} \text{ s}^{-1}\)) & \(\Omega\) (\(\times 10^{3} \text{ s}^{-1}\)) & \(\gamma / \Omega\) \\
\hline\hline
NVA1 & 23.9(6) & n/a & n/a & n/a & 0.13(2) & 0.063(9) & 2.1(4) \\
NVA1 & 125.9(6) & n/a & n/a & n/a & 0.111(9) & 0.053(3) & 2.1(2) \\
NVA1 & 233.2(6) & n/a & n/a & n/a & 0.132(17) & 0.061(6) & 2.2(4) \\
\hline
NVA2 & 129.7(6) & n/a & n/a & n/a & 0.114(12) & 0.060(4) & 1.9(2) \\
\hline
NVA3 & 121.2(6) & 32.1 & 21.6 & 23.8 & 0.133(10) & 0.050(4) & 2.7(3) \\
NVA3 & 120.8(6) & 21.6 & 21.6 & 0.6 & 0.114(10) & 0.059(4) & 1.9(2) \\
\hline
NVB1 & 167.1(6) & n/a & n/a & n/a & 0.132(11) & 0.056(3) & 2.4(2) \\
NVB1 & 831.6(6) & n/a & n/a & n/a & 0.17(4) & 0.08(2) & 2.0(7) \\
NVB1 & 1207.1(6) & n/a & n/a & n/a & 0.10(3) & 0.048(13) & 2.1(8) \\
NVB1 & 40.6(6) & n/a & n/a & n/a & 0.12(1) & 0.049(3) & 2.3(2) \\
NVB1 & 412.7(6) & n/a & n/a & n/a & 0.135(13) & 0.065(5) & 2.1(3) \\
NVB1 & 623.8(6) & n/a & n/a & n/a & 0.138(14) & 0.061(5) & 2.3(3) \\
NVB1 & 187.2(6) & 33.4 & 33.4 & 0.0 & 0.121(10) & 0.062(5) & 2.0(2) \\
NVB1 & 80.4(6) & 32.5 & 14.4 & 29.2 & 0.145(11) & 0.06(5) & 2.4(3) \\
NVB1 & 177.2(6) & 33.0 & 31.6 & 9.3 & 0.130(10) & 0.058(5) & 2.2(3) \\
NVB1 & 121.4(6) & 32.1 & 21.7 & 23.7 & 0.158(12) & 0.052(4) & 3.0(3) \\
NVB1 & 33.7(6) & 32.9 & 6.0 & 32.4 & 0.139(13) & 0.059(5) & 2.4(3) \\
NVB1 & 154.5(6) & 32.6 & 27.6 & 17.3 & 0.135(11) & 0.056(5) & 2.4(3) \\
NVB1 & 122.1(6) & 32.7 & 21.8 & 24.4 & 0.150(12) & 0.055(5) & 2.7(3) \\
NVB1 & 122.7(6) & 21.9 & 21.9 & 0.0 & 0.113(10) & 0.055(4) & 2.1(2) \\
NVB1 & 16.6(6) & 32.2 & 2.9 & 32.1 & 0.128(19) & 0.054(5) & 2.4(4) \\
NVB1 & 555.7(6) & 146.5 & 98.8 & 108.1 & 0.16(2) & 0.064(9) & 2.5(5) \\
NVB1 & 1662.4(6) & 300.0 & 297.5 & 38.7 & 0.149(15) & 0.053(7) & 2.8(5) \\
\hline
NVE & 303.4(6) & 55.1 & 54.0 & 10.6 & 0.140(12) & 0.047(3) & 3.0(3) \\
NVE & 191.1(6) & 54.9 & 34.1 & 43.1 & 0.147(14) & 0.065(5) & 2.3(3) \\
NVE & 261.3(6) & 55.0 & 46.6 & 29.3 & 0.149(13) & 0.067(6) & 2.2(3) \\
NVE & 130.4(6) & 56.1 & 23.7 & 50.8 & 0.185(17) & 0.057(5) & 3.2(4) \\
NVE & 25.3(6) & 53.4 & 4.3 & 53.2 & 0.24(3) & 0.064(7) & 3.7(6) \\
NVE & 324.3(6) & 57.8 & 57.8 & 0.0 & 0.120(11) & 0.061(4) & 2.0(2) \\
NVE & 294.0(6) & 58.6 & 52.5 & 26.1 & 0.151(15) & 0.065(5) & 2.3(3) \\
NVE & 29.4(6) & 54.5 & 4.8 & 54.3 & 0.20(3) & 0.070(6) & 2.8(4) \\
NVE & 251.9(6) & 60.9 & 45.1 & 41.0 & 0.140(14) & 0.052(4) & 2.7(3) \\
NVE & 217.4(6) & 60.0 & 38.9 & 45.8 & 0.154(15) & 0.064(5) & 2.4(3) \\
NVE & 40.3(6) & 53.7 & 7.1 & 53.3 & 0.181(19) & 0.055(4) & 3.3(4)
\end{tabular}
\end{table*}

\clearpage
\bibliography{SuppReferences.bib}